\documentclass[11pt,a4paper]{article}
\usepackage[hyperref]{acl2019}
\usepackage{times}
\usepackage{latexsym}
\usepackage{latexsym}
\usepackage{graphicx}
\usepackage{multirow}
\usepackage[toc,page]{appendix}
\usepackage{amsmath}
\usepackage{listings}
\usepackage{comment}
\usepackage{tabularx}
\usepackage{ragged2e}
\usepackage{pgfplots}

\usepackage{url}

\usepackage{dramatist}

\aclfinalcopy 
\def\aclpaperid{23} 

\title{Affective Behaviour Analysis of On-line User Interactions:\\Are On-line Support Groups more Therapeutic than Twitter?}

\author{Giuliano Tortoreto$^\dagger$, Evgeny A. Stepanov$^\dagger$, Alessandra Cervone$^\dagger$, \\ \textbf{Mateusz Dubiel}$^\star$, \textbf{Giuseppe Riccardi}$^\dagger$ \\
$^\dagger$Signals and Interactive Systems Lab, University of Trento, Italy\\
$^\star$University of Strathclyde, Glasgow, UK\\
{\tt name.surname@unitn.it, name.surname@strath.ac.uk}}

\date{}

\usepackage{url}
\newcommand{\online}{on-line }

\newcommand{\osg}{on-line support groups }
\newcommand{\osgn}{on-line support groups}
\newcommand{\osgacronymn}{OSG}
\newcommand{\osgacronym}{OSG }

\usepackage{xparse}

\newsavebox{\fminipagebox}
\NewDocumentEnvironment{fminipage}{m O{\fboxsep}}
 {\par\kern#2\noindent\begin{lrbox}{\fminipagebox}
  \begin{minipage}{#1}\ignorespaces}
 {\end{minipage}\end{lrbox}%
  \makebox[#1]{%
    \kern\dimexpr-\fboxsep-\fboxrule\relax
    \fbox{\usebox{\fminipagebox}}%
    \kern\dimexpr-\fboxsep-\fboxrule\relax
  }\par\kern#2
 }

\usepackage{graphicx}

\newcolumntype{R}{>{\raggedleft\arraybackslash}X}

\graphicspath{{images/}}

\begin{document}

\maketitle


\begin{abstract}
The increase in the prevalence of mental health problems has coincided with a growing popularity of health related social networking sites. 
Regardless of their therapeutic potential, \osg (OSGs) can also have negative effects on patients. In this work we propose a novel methodology to automatically verify the presence of therapeutic factors in social networking websites by using Natural Language Processing (NLP) techniques.
The methodology is evaluated on \online asynchronous multi-party conversations collected from an \osgacronym and Twitter.
The results of the analysis indicate that therapeutic factors occur more frequently in \osgacronym conversations than in Twitter conversations.
Moreover, the analysis of \osgacronym conversations reveals that the users of that platform are supportive, and interactions are likely to lead to the improvement of their emotional state. 
We believe that our method provides a stepping stone towards automatic analysis of emotional states of users of online platforms. Possible applications of the method  include provision of guidelines that highlight potential implications of using such platforms on users' mental health, and/or support in the analysis of their impact on specific individuals.


\end{abstract}

\DeclareGraphicsExtensions{.pdf,.png,.jpg}

\section{Introduction}
Recently, people have started looking at online forums either as a primary or secondary source of counseling services \cite{vogel2007avoidance}. \citet{reachoutanalysis} reported that over the first five years of operation (2011-2016), ReachOut.com -- Ireland's online youth mental health service -- 62\% of young people would visit a website for support when going through a tough time. With the expansion of the Internet, there has been a substantial growth in the number of users looking for psychological support online. 

The importance of the \online life of patients has been recognized in research as well. 
\citet{AmichaiHamburger2014288} stated that the online life of patients constitutes a major influence on their self-definition. Furthermore, according to \citet{back2010facebook}, the social networking activities of an individual, offer an important reflection of their personality. While dealing with patients suffering from psychological problems, it is important that therapists do not ignore this pivotal source of information which can provide deep insights into their patients' mental conditions.

Acceptance of \osg (\osgacronymn) by Mental Health Professionals is still not established \cite{ecp2017gerhardMaximizing}. Since \osgacronym can have double-edged effects on patients and the presence of professionals is often limited, we argue that their properties should be further studied.
According to \citet{barak2008fostering} \osgacronym effectiveness is hard to assess, while some studies showed \osgacronymn's potential to change participants' attitudes, no such effect was observed in other studies (see Related Work Section for more details).
Furthermore the scope of previous work on analysis of users' behaviour in \osgacronym has been limited by the fact that they relied on expert annotation of posts and comments \cite{MayfieldDiscoveringHabits}.

We present a novel approach for automatically analysing online conversations for the presence of therapeutic factors of group therapy defined by \citet{yalom2005theory} as ``the actual mechanisms of effecting change in the patient''.
The authors have identified 11 therapeutic factors in group therapy: Universality, Altruism, Instillation of Hope, Guidance, Imparting information, Developing social skills, Interpersonal learning, Cohesion, Catharsis, Existential factors, Imitative behavior and Corrective recapitulation of family of origin issues.
In this paper, we focus on 3 therapeutic factors: Universality, Altruism and Instillation of Hope (listed below), as we believe that these can be  approximated by using established NLP techniques (e.g. Sentiment Analysis, Dialogue Act tagging etc.).

\begin{enumerate}
\itemsep0em
	\item Universality: the disconfirmation of a user's feelings of uniqueness of their mental health condition.
	\item Altruism: others offer support, reassurance, suggestions and insight.
	\item Instillation of Hope: inspiration provided to participants by their peers.
\end{enumerate}

The selected therapeutic factors are analysed in terms of illocutionary force\footnote{The illocutionary force of an utterance is the speaker's intention in producing that utterance according to \citet{glossarylinguisticloos}.} and attitude\footnote{``The attitude may be either his or her affective state, namely the emotional state of the author when writing, or the intended emotional communication, namely the emotional effect the author wishes to have on the reader'' \citet{Gala:2014:LPC:2695511}.}.
Due to the multi-party and asynchronous nature of \online social media conversations, prior to the analysis, we extract conversation threads among users -- an essential prerequisite for any kind of higher-level dialogue analysis \cite{Elsner:2010:DC:1950488.1950492}. Afterwards, the illocutionary force is identified using Dialogue Act tagging, whereas the attitude by using Sentiment Analysis. The quantitative analysis is then performed on these processed conversations. 

Ideally, the analysis would require experts to annotate each post and comment on the presence of therapeutic factors. However, due to time and cost demands of this task, it is feasible to analyse only a small fraction of the available data. 
Compared to previous studies (e.g. \cite{MayfieldDiscoveringHabits}) that analysed few tens of conversations and several thousand lines of chat; using the proposed approach -- application of Dialogue Acts and Sentiment Analysis -- we were able to automatically analyse approximately 300 thousands conversations (roughly 1.5 million comments). 

The rest of the paper is structured as follows. 
In Section 2 we introduce related work. 
Next, in Section 3 we describe the pre-processing pipeline and the methodology to perform thread extraction on asynchronous multi-party conversations.
In Section 4 we provide the describe the final dataset used for the analysis, and in Section 5 we present the results of our analysis. 
Finally, in Section 6 we provide concluding remarks and future research directions.

\section{Related Work} 
On-line support groups have been analyzed for various factors before. For instance, \citet{osgsBenefitsPatientsChung} analysed stress reduction in \online support group chat-rooms, and the effects of \online social interactions. 
Such studies mostly relied on questionnaires and were based on a small number of users. Nevertheless, in \citet{osgsBenefitsPatientsChung}, the author showed that social support facilitates coping with distress, improves mood and expedites recovery from it. These findings highlight that, overall, \online discussion boards appear to be therapeutic and constructive for individuals suffering alcohol-abuse.

Application of NLP to the analysis of mental health-related conversation has been studied as well (e.g. \cite{ghosh2017you,stepanov2018depression}).
\citet{MayfieldDiscoveringHabits} applied sentiment-analysis combined with extensive turn-level annotation to investigate stress reduction in \online support group chat-rooms, showing that sentiment-analysis is a good predictor of entrance stress level. Furthermore, similar to our setting, they applied automatic thread-extraction to determine conversation threads.

\citet{kissane2007supportive} have shown that \online support group therapy increased the quality of life of patients with metastatic breast cancer. Since many original posters reported the benefits of group therapy on patients \cite{effgroupthe,AmichaiHamburger2014288,bengroupthe,espie2012randomized,gary2000online,yalom2005theory}, we evaluate the effect of the user interaction using sentiment scores of comments in \osgn.

According to \citet{MayfieldDiscoveringHabits}, users with high incoming stress tend to request less information from others, as a percentage of their time, and share much more information, in absolute terms. In addition, high information sharing has been shown to be a good predictor of stress reduction at the end of the chat \cite{MayfieldDiscoveringHabits}.
Regarding information sharing, we rely on Dialogue Acts \cite{austin1975things} to model the speaker's intention in producing an utterance. In particular, we are interested in Dialogue Act label that is defined to represent descriptive, narrative, or personal information -- the \textit{statement}.

Dialogue Acts have been applied to the analysis of spoken \cite{Stolcke:2000:DAM:971869.971872,cervone2018automatically} as well as \online written synchronous conversations \cite{forsythand2007lexical}.
We apply Dialogue Act tag set defined in \citet{forsythand2007lexical} to the analysis of our \online asynchronous conversations. 
We argue that Dialogue Acts can be used to analyse user behaviour in social media and verify the presence of therapeutic factors.



\section{Methodology}
We select the three therapeutic factors -- Universality, Altruism and Instillation of Hope -- that can be best approximated using NLP techniques: Sentiment Analysis and Dialogue Act tagging.
We discuss each one of the selected therapeutic factors and the identified necessary conditions. 
The listed conditions, however, are not sufficient to attribute the presence of a therapeutic factor with high confidence, which only can be obtained using expert annotation. 
Our analysis focuses on the structure of conversations; though content plays an important role as well. 
 
{\em Universality} consists in the disconfirmation of patients' belief of uniqueness of their disease. This therapeutic factor is shown to be a powerful source of relief for the patient, according to \citet{yalom2005theory}. From this definition, we can draw the following conditions that are applicable to our environment:
\begin{enumerate}
\itemsep0em
	\item improvement of original poster's sentiment: we hypothesize that the discovery that other people passed through similar issues leads to a higher sentiment score;
	\item posts containing negative personal experiences: to disconfirm the belief of uniqueness users have to share their story;
	\item comments containing negative statements: to disconfirm the  patient's feelings of uniqueness, the commenting user must tell a similar negative personal experience. This condition requires two sub-conditions: high presence of statements in comments and the presence of negative comments replying to negative posts.
\end{enumerate}

{\em Instillation of Hope} is based on inspiration provided to participants by their peers. Through the inspiration provided by their peers, patients can increase their expectation on the therapy outcome. \citet{yalom2005theory} in several studies have demonstrated that a high expectation of help before the start of a therapy is significantly correlated with a positive therapy outcome. The author states that many patients pointed out the importance of having observed the improvement of others. Therefore, the three main conditions are the following:
\begin{enumerate}
\itemsep0em
	\item improvement of original poster's sentiment: we hypothesize that instillation of hope leads to a higher sentiment score;
	\item posts containing negative personal experiences: hope can be instilled in someone who shares a negative personal experience;
	\item comments containing positive personal experiences: in order to instill hope, commenting posters must show to original posters an overall positive personal experience. To detect positive personal experience, we require the presence of statements in comments and a positive sentiment of comments replying to negative posts.
\end{enumerate}

{\em Altruism} consists of peers offering support, reassurance, suggestions and insight, since they share similar problems with one another \cite{yalom2005theory}. The experience of finding that a patient can be of value to others is refreshing and boosts self-esteem \cite{yalom2005theory}. However, in the current study we focus on testing whether commenting posters are altruists or not. We do not test whether the altruistic behavior leads to an improvement on the altruist itself. For these reasons, we define three main conditions:
\begin{enumerate}
\itemsep0em
	\item improvement of original poster's sentiment: we hypothesize that supportive and reassuring statements improve the sentiment score of the original poster;
	\item posts contains negative personal experiences: users offer support, reassurance and suggestion when facing a negative personal experience of the original poster;
	\item comments containing positive statements: either supportive or reassuring statements show by definition a positive intended emotional communication. Thus comments to the post should consist of positive sentiment statements.
\end{enumerate}

Consequently, a conversation containing the aforementioned therapeutic factors should satisfy the following conditions in terms of NLP: Sentiment Analysis and Dialogue Acts.

\begin{enumerate}
\itemsep0em
	\item original posters have a higher sentiment score at the end of the thread than at the beginning;
	\item the original post consists mostly of polarised statements;
	\item the presence of a significant amount of statements in comments, since both support and sharing similar negative experiences can be represented as statements;
	\item both negative and positive statements in comments lead to higher final sentiment score of the original poster.
\end{enumerate}


\section{Datasets}
\begin{figure*}
\begin{fminipage}{\textwidth}
\begin{drama}
  \Character{Alice}{alice}
  \Character{Bob}{bob}

  \alicespeaks: I want to tell him that if he can't have a real conversation with me then don't talk to me, because it hurts more to feel like I'm an obligation.... I don't want anyone to ever get close to me but I don't want to be alone.

  \bobspeaks to feel like an obligation is really disheartening and takes a stab at the self esteem. What about the conversations make you feel like a an obligation? Have you talked to him about this?

  \alicespeaks He doesn't have a conversation, I know he doesn't mean to, he's just always busy now... I don't want to make him feel bad.
  
  \bobspeaks Just remember that your needs matter too!
  
  ...
  
  \alicespeaks @Bob Thank you :)
\end{drama}
\end{fminipage}
\caption{An example of conversation thread extracted from \osgacronymn. }
 \label{fig:conv_example}
\end{figure*}
\begin{figure*}
\begin{fminipage}{\textwidth}
\begin{drama}
  \Character{Carol}{carol}
  \Character{Dave}{dave}

  \carolspeaks: lol my best friend at the time got cheated on, we wrote on the guys truck..he started chasing us, i tripped and broke my ankle \#justmyluck

  \davespeaks wtf

  \carolspeaks it was ridiculous  and i drove with my foot hanging out the window all f***ed up
  
  \davespeaks  when was this i'm so confused 

\end{drama}
\end{fminipage}
\caption{An example conversation thread extracted from Twitter. }
 \label{fig:conv_example}
\end{figure*}

We verify the presence of therapeutic factors in two social media datasets: OSG and Twitter. The first dataset is crawled from an \osg website, and the second dataset consists of a small sample of Twitter conversation threads.
Since the former consists of multi-threaded conversations, we apply a pre-processing to extract conversation threads to provide a fair comparison with the Twitter dataset. An example conversation from each data source is presented in Figure \ref{fig:conv_example}.

\subsection{Twitter}
We have downloaded 1,873 Twitter conversation threads, roughly 14k tweets, from a publicly available resource\footnote{https://github.com/Phylliida/Dialogue-Datasets} that were previously pre-processed and have conversation threads extracted. A conversation in the dataset consists of at least 4 tweets.
Even though, according to \citet{paul2011you}, Twitter is broadly applicable to public health research, our expectation is that it contains less therapeutic conversations in comparison to specialized \online support forums.

\subsection{OSG}
Our data has been developed by crawling and pre-processing an \osgacronym web forum. The forum has a great variety of different groups such as depression, anxiety, stress, relationship, cancer, sexually transmitted diseases, etc. Each conversation starts with one post and can contain multiple comments.  Each post or comment is represented by a poster, a timestamp, a list of users it is referencing to, thread id, a comment id and a conversation id. The thread id is the same for comments replying to each other, otherwise it is different. The thread id is increasing with time. Thus, it provides ordering among threads; whereas the timestamp provides ordering in the thread.

Each conversation can belong to multiple groups. Consequently, the dataset needs to be processed to remove duplicates. The dataset resulting after de-duplication contains 295 thousand conversations, each conversation contains on average 6 comments. In total, there are 1.5 million comments. 
Since the created dataset is multi-threaded, we need to extract conversation threads, to eliminate paths not relevant to the original post. 

\subsubsection{Conversation Thread Extraction}
The thread extraction algorithm is heuristic-based and consists of two steps:
(1) creation of a tree, based on a post written by a user and the related comments and
(2) transformation of the tree into a list of threads.

The tree creation is an extension of the approach of \citet{Gomez:2008:SAS:1367497.1367585}, where first a graph of conversation is constructed. In the approach, direct replies to a post are attached to the first nesting level and subsequent comments to increasing nesting levels. In our approach, we also exploit comments' features.

The tree creation is performed without processing the content of comments, which allows us to process posts and comments of any length efficiently.
The heuristic used in the process is based on three simplifying assumptions:
\begin{enumerate}
\itemsep0em
\item Unless there is a specific reference to another comment or a user, comments are attached to the original post.
\item When replying, the commenting poster is always replying to the original post or some other comment. Unless specified otherwise, it is assumed that it is a response to the previous (in time) post/comment.
\item Subsequent comments by the same poster are part of the same thread.
\end{enumerate}




To evaluate the performance of the thread extraction algorithm, 2 annotators have manually constructed the trees for 100 conversations. 
The performance of the algorithm on this set of 100 conversations is evaluated using accuracy and standard Information Retrieval evaluation metrics of precision, recall, and F$_1$ measure. The results are reported in Table \ref{table:Comparison results} together with random and majority baselines. 
The turn-level percent agreement between the 2 annotators is 97.99\% and Cohen's Kappa Coefficient is 83.80\%.

\begin{table}[t]
\centering
\begin{tabular}{|l|l|l|l|l|}
\hline
\textbf{Approach}     & \textbf{Acc} & \textbf{P} & \textbf{R} & \textbf{F$_1$} \\ \hline
Majority Baseline     & 0.92              & 0.46               & 0.46            & 0.46              \\
Random Baseline       & 0.87              & 0.14               & 0.14            & 0.14              \\ \hline
\textbf{Our Approach} & \textbf{0.97}     & \textbf{0.79}      & \textbf{0.80}   & \textbf{0.80}     \\
\hline
\end{tabular}
\caption{\label{table:Comparison results} Performance of the thread extraction algorithm on a set of 100 manually constructed trees.}
\label{eval_res_100}
\end{table}

\subsection{Data Representation}
For both data sources, Twitter and OSG with extracted threads, posts and comments are tokenized\footnote{NLTK sentence tokenizer} and sentence split.
Each sentence is passed through Sentiment Analysis and Dialogue Act tagging. 
Since a post or a comment can contain multiple sentences, therefore multiple Dialogue Acts, it is represented as as a one-hot encoding, where each position represents a Dialogue Act.

For Sentiment Analysis we use a lexicon-based sentiment analyser introduced by \citet{alistair2005sentiment}. For Dialogue Act tagging, on the other hand, we make use of a model trained on NPSChat corpus \cite{forsythand2007lexical} following the approach of  \citet{lan2008dialogue}.\footnote{The model achieves 80.21\% accuracy.} 


\section{Analysis}
\label{sec:analysis}
As we mentioned in Section 3, the presence of each of the therapeutic conditions under analysis is a necessary for a conversation to be considered to have therapeutic factors. In this section we present the results of our analysis with respect to these conditions.

\subsection{Change in Sentiment score of Original Posters}
\begin{figure}
    \centering
\includegraphics[width=7.0cm]{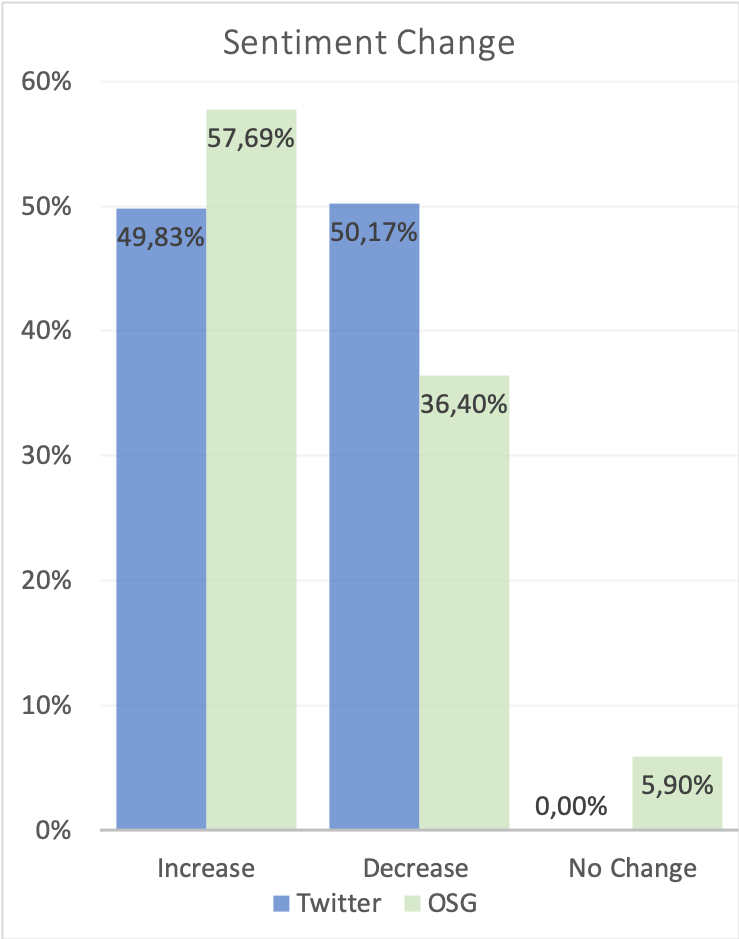}
      \caption{The percentages of threads in \osgacronym and Twitter leading to the increase or decrease of the sentiment score of the original poster.}

 \label{fig:overall_increase}
\end{figure}

\begin{figure}[t]
    \centering
\includegraphics[width=0.75\columnwidth]{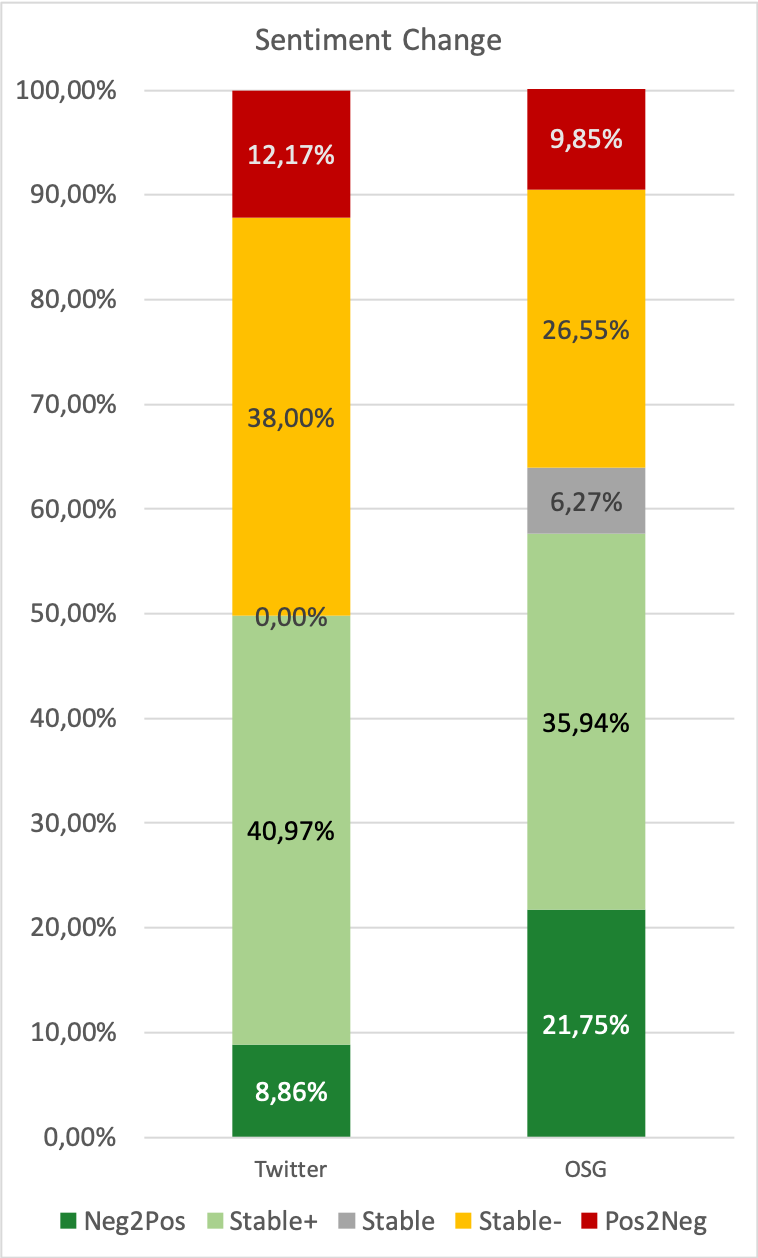}
\caption{The sentiment polarity change in the two datasets - Twitter and \osgacronymn. 
Stable segments are labeled either as an increase (+), decrease (-) or no change in polarity, including neutral comments. Pos2Neg and Neg2Pos denote a nominal polarity change.}
 \label{fig:cumul_overall}
\end{figure}

The first condition which we test is the sentiment change in conversation threads, comparing the initial and final sentiment scores (i.e. posts' scores) of the original poster.
The results of the analysis are presented in Figure \ref{fig:overall_increase}. In the figure we can observe that the distribution of the sentiment change in the two datasets is different. While in Twitter the amount of conversations that lead to the increase of sentiment score is roughly equal to the amount of conversations that lead to the decrease of sentiment score; the situation is different for \osgacronymn. In \osgacronymn, the amount of conversations that lead to the increase of sentiment score is considerably higher.


Figure \ref{fig:cumul_overall} provides a more fine grained analysis, where we additionally analyse the sentiment change in nominal polarity terms -- negative and positive.
In \osgacronymn, the number of users that changed polarity from negative to positive is more than the double of the users that have changed the polarity from positive to negative.
In Twitter, on the other hand, the users mostly changed polarity from positive to negative.  Results of the analysis suggest that in \osgacronym, sentiment increases and users tend to change polarity from negative to positive, whereas in Twitter sentiment tends to decrease.
Verification of this condition alone indicates that the ratio of potentially therapeutic conversations in Twitter is lower.

\subsection{Structure of Posts and Comments}

\begin{table}
\centering
\begin{tabular}{| l | r | r |}
\hline
\textbf{Class} &  \textbf{Twitter} & \textbf{\osgacronymn} \\
\hline
Statement & 62.9 & 73.0 \\
Emphasis & 9.6 & 6.3 \\
ynQuestion & 7.5 & 4.7 \\   
Continuer & 2.5 & 4.3 \\
whQuestion & 6.1 & 3.7 \\
Reject & 2.6 & 2.9 \\
Emotion & 2.9 & 1.5 \\
Accept & 2.4 & 1.3 \\
Greet & 0.6 & 0.8 \\
nAnswer & 1.1 & 0.4 \\
yAnswer &  0.8 & 0.3\\
Bye & 0.4 & 0.2 \\
Clarify & 0.1 & $< 0.1$ \\
Other & $< 0.1$ & $< 0.1$ \\
\hline
\end{tabular}
\caption{The distribution (in percentages) of automatically predicted per-sentence Dialogue Act tags. Tags are counted separately for each sentence in the multi-sentence posts and comments.}
\label{table:predicted_classes}
\end{table}

Table \ref{table:predicted_classes} presents the distribution of automatically predicted per-sentence Dialogue Acts in the datasets. The most frequent tag is \textit{statement} in both. 
In Table \ref{tab:da_dist}, on the other hand, we present the distribution of post and comment structures in terms of automatically predicted Dialogue Act tags. The structure is an unordered set of tags in the post or comment. 
From the table we can observe that the distribution of tag sets is similar between posts and comments. In both cases the most common set is \textit{statement} only. However, conversations containing only \textit{statement}, \textit{emphasis} or \textit{question} posts and comments predominantly appear in Twitter. Which is expected due to the shorter length of Twitter posts and comments.

\begin{table}[h]
\centering
\begin{tabularx}{\columnwidth}{|l|R|R|R|R|}
\hline
\textbf{Tag Set} 
& \multicolumn{2}{c|}{\textbf{Posts}}
& \multicolumn{2}{c|}{\textbf{Comments}} \\
& Twitter & \osgacronym
& Twitter & \osgacronym \\
\hline
Statement  & 64.12 & 38.79 & 57.14 & 41.45 \\
Emphasis   &   3.01 &  1.31 &  4.42 &  3.96 \\
ynQuestion &   4.79 &  2.94 &  4.80 &  2.14 \\
whQuestion &   4.00 &  1.43 &  4.86 &  2.07 \\
\hline\hline
\multicolumn{5}{|c|}{Statement +}\\
\hline
Emphasis   & 2.17 & 3.96 & 3.65 & 5.57 \\
Continuer  & 0.99 & 6.29 & 0.92 & 4.59 \\
ynQuestion & 2.86 & 7.04 & 1.92 & 4.05 \\
whQuestion & 4.00 & 3.98 & 1.56 & 2.95 \\
Accept     & 0.44 & 0.81 & 0.19 & 1.92 \\
Reject     & 1.28 & 3.00 & 0.95 & 3.38 \\
\hline
\end{tabularx}
    \caption{The distribution (in percentages) of post and comment structures represented as unordered set of Dialogue Act tags.}
    \label{tab:da_dist}
\end{table}

We can also observe that the original posters tend to ask more questions than the commenting posters -- 19.83\% for posts vs. 11.21\% for comments (summed).
This suggests that the original posters frequently ask either for suggestion or confirmation of their points of view or their disconfirmation. 
However, the high presence of personal experiences is supported by the high number of posts containing only statements. 

High number of \textit{statement} tags in comments suggests that users reply either with supporting or empathic statements or personal experience. However, 6.39\% of comments contain \textit{accept} and \textit{reject} tags, which mark the degree to which a speaker accepts some previous proposal, plan, opinion, or statement \cite{Stolcke:2000:DAM:971869.971872}. The described Dialogue Act tags are often used when commenting posters discuss original poster's point of view. For instance,
``It's true. I felt the same.'' -- \textit{\{Accept, Statement\}} or ``Well no. You're not alone'' -- \textit{\{Reject, Statement\}}.
The datasets differ with respect to the distribution of these Dialogue Acts tags, they appear more frequently in \osgacronymn.



\subsection{Sentiment of Posts and Comments} 
\begin{table}[h]
\centering
\begin{tabular}{|l|c|c|c|}
\hline
& \multicolumn{3}{c|}{\textbf{Sentiment Polarity}}\\
& Negative & Neutral & Positive\\
\hline\hline
\multicolumn{4}{|c|}{OSG}\\
\hline
Posts    & 32.1 & 33.5 & 34.5 \\
Comments & 25.8 & 31.7 & 42.5 \\
\hline\hline
\multicolumn{4}{|c|}{Twitter}\\
\hline
Posts    & 20.5 & 44.0 & 35.5 \\
Comments & 21.1 & 45.9 & 33.0 \\
\hline

\end{tabular}
\caption{The distribution (in percentages) of sentiment in \textit{statement} sentences of posts and comments.}
\label{table:statement_sentiment}
\end{table}

Table \ref{table:statement_sentiment} presents the distribution of sentiment polarity in post and comment statements (i.e. sentences tagged as \textit{statement}). For \osgacronymn, the predominant sentiment label of statements is positive and it is the highest for both posts and comments.
However, the difference between the amounts of positive and negative statements is higher for the replying comments (34.5\% vs. 42.5\%).
For Twitter, on the other hand, the predominant sentiment label of statements is neutral and the polarity distribution between posts and comments is very close. One particular observation is that the ratio of negative statements is higher in \osgacronym for both posts and comments than in Twitter, which supports the idea of sharing negative experiences.

Further we analyze whether the sentiment of a comment (i.e. the replying user) is affected by the sentiment of the original post (i.e. the user being replied to), which will imply that the users adapt their behaviour with respect to the post's sentiment. 
For the analysis, we split the datasets into three buckets according to the posts' sentiment score -- negative, neutral, or positive, and represent each conversation in terms of percentages of comments (replies) with each sentiment label. The buckets are then compared using t-test for statistically significant differences.

\begin{table}[h]
\centering
\begin{tabular}{|l|c|c|c|}
\hline
\textbf{Posts} & \multicolumn{3}{c|}{\textbf{Comments}}\\
    & Negative & Neutral & Positive \\ 
\hline\hline
\multicolumn{4}{|c|}{OSG}\\
\hline
Negative & 27.25 & 14.87 & 57.88 \\
Neutral  & 21.37 & 23.49 & 55.14 \\
Positive & 22.79 & 19.62 & 65.17 \\
\hline\hline
\multicolumn{4}{|c|}{Twitter}\\
\hline
Negative & 32.92 & 22.85 & 44.23 \\
Neutral  & 26.48 & 25.00 & 48.52 \\
Positive & 18.79 & 16.04 & 57.60 \\
\hline
\end{tabular}
\caption {The distribution (in percentages) of reply sentiment labels with respect to the post's sentiment label.}
\label{table:reply_sentiment}
\end{table}

Table \ref{table:reply_sentiment} presents the distribution of sentiment labels with respect to the post's sentiment score. The patterns of distribution are similar across the datasets. We can observe that overall, replies tend to have a positive sentiment, which suggests that replying posters tend to have a positive attitude. However, the ratio of positive comments is higher for \osgacronym than for Twitter.

The results of the Welch's t-test on \osgacronym data reveal that there are statistically significant differences in the distribution of replying comments' sentiment between conversations with positive and negative starting posts. A positive post tends to get significantly more positive replies. Similarly, a negative post tends to get significantly more negative replies (both with $p < 0.01$).

\begin{table}[h]
\centering
\begin{tabular}{|l|c|c|c|}
\hline
\textbf{Comments} & \multicolumn{3}{c|}{\textbf{Final Sentiment of OP}}\\
    & Negative & Neutral & Positive \\ 
\hline\hline
\multicolumn{4}{|c|}{OSG}\\
\hline
Negative & 28.98 & 18.27 & 52.75 \\
Neutral  & 25.20 & 25.70 & 49.10 \\
Positive & 22.60 & 18.01 & 59.38 \\
\hline\hline
\multicolumn{4}{|c|}{Twitter}\\
\hline
Negative & 24.14 & 41.78 & 34.08\\
Neutral  & 21.34 & 49.16 & 29.50 \\
Positive & 20.25 & 35.20 & 44.55 \\
\hline
\end{tabular}
\caption {The distribution (in percentages) of sentiment labels of the final text of the original poster (OP) with respect to the comment's sentiment label.}
\label{table:react_sentiment}
\end{table}

Table \ref{table:react_sentiment} presents the distribution of the sentiment labels of the final text provided by the original poster with respect to the sentiment polarity of the comments. The results indicate that OSG participants are more supportive, as the majority of conversations end in a positive final sentiment regardless of the sentiment of comments. We can also observe that negative comments in OSG lead to positive sentiment, which supports the idea of sharing the negative experiences, thus presence of therapeutic factors.
For Twitter, on the other hand, only positive comments lead to the positive final sentiments, whereas other comments lead predominantly to neutral final sentiments.

Our analysis in terms of sentiment and Dialogue Acts supports the presence of the three selected therapeutic factors -- Universality, Altruism and Instillation of Hope -- in \osgacronym more than in Twitter. The main contributors to this conclusion are the facts that there is more positive change in the sentiment of the original posters in OSG (people seeking support) and that in OSG even negative and neutral comments are likely to lead to positive changes.

\section{Conclusion}
In this work, we propose a methodology to automatically analyse online social platforms for the presence of therapeutic factors (i.e. Universality, Altruism and Instillation of Hope). We evaluate our approach on two \online platforms, Twitter and an \osgacronym web forum. We apply NLP techniques of Sentiment Analysis and Dialogue Act tagging to automatically verify the presence of therapeutic factors, which allows us to analyse larger amounts of conversational data (as compared to previous studies). 

Our analysis indicates that \osgacronym conversations satisfy higher number of conditions approximating therapeutic factors than Twitter conversations. Given this outcome, we postulate that users who join support group websites spontaneously seem to benefit from it. Indeed, as shown in Section \ref{sec:analysis}, the original posters who interact with others by replying to comments, have benefited from an improvement of their emotional state.

We would like to reemphasise that the conditions for the therapeutic factors are necessary but not sufficient; since our analysis focuses on the structure of conversations, being agnostic to the content. NLP, however, allows us to strengthen our approximations even further. 
Thus, the further extension of our work is also augmentation of our study with other language analysis metrics and their correlation with human annotation.

It should be noted that the proposed approach is an approximation of the tedious tasks of annotation of conversations by experts versed in the therapeutic factors and their associated theories. Even though we can use Sentiment Analysis to detect the existence of therapeutic factors, we cannot differentiate between Altruism and Instillation of Hope, as this requires differentiation between emotional state of the user and the intended emotional communication. Thus, the natural extensions of this work are differentiation between different therapeutic factors and comparison of the proposed analysis to the human evaluation.

Although we acknowledge that the proposed methodology does not serve as a replacement of manual analysis of \osgacronym for the presence of therapeutic factors, we believe that it could facilitate and supplement this process. The method can serve as a tool for general practitioners and psychologists who can use it as an additional source of information regarding their patients’ condition and, in turn, offer a more personalised support that is better tailored to individual therapeutic needs.

\newpage
\bibliography{acl2019}
\bibliographystyle{acl_natbib}
\balance

\end{document}